\documentclass[11pt]{article}

\input epsf

\begingroup\makeatletter\ifx\SetFigFont\undefined%
\gdef\SetFigFont#1#2#3#4#5{%
  \reset@font\fontsize{#1}{#2pt}%
  \fontfamily{#3}\fontseries{#4}\fontshape{#5}%
  \selectfont}%
\fi\endgroup%

\newcommand{\nc}{\newcommand}

\nc{\eqr}[1]{(\ref{#1})}
\nc{\sref}[1]{\S~\ref{#1}}
\nc{\beq}{\begin{equation}}
\nc{\eeq}{\end{equation}}
\nc{\barray}{\begin{eqnarray}}
\nc{\earray}{\end{eqnarray}}
\nc{\barrayn}{\begin{eqnarray*}}
\nc{\earrayn}{\end{eqnarray*}}
\nc{\bcenter}{\begin{center}}
\nc{\ecenter}{\end{center}}

\def\gd{\delta}

\def\ve{\varepsilon}

\def\gk{\kappa}

\def\ksM{km sec$^{-1}$ Mpc$^{-1}$}
\def\Om{\Omega}
\def\Lam{\Lambda}
\def\dd#1#2{{d#1\over d#2}}

\newcommand{\pone}[1]{\left( #1 \right)}
\newcommand{\ptwo}[1]{\left[ #1 \right]}
\newcommand{\pthree}[1]{\left\{ #1 \right\}}
\newcommand{\ev}[1]{\left< #1 \right>}

\renewcommand{\thefootnote}{\fnsymbol{footnote}}

\begin{document}
\begin{titlepage}


\vspace*{1cm}
\begin{center}
{\LARGE A Simple Quantum Cosmology}
\end{center}

\vspace{1cm}
\begin{center}
{\sc T. R. Mongan \footnote{E-mail: tmongan@mail.com}} \\
{\it 84 Marin Avenue\\
Sausalito, CA 94965\\ U.S.A.}
\end{center}

\vspace{1cm}

\begin{abstract}
\noindent

 A simple and surprisingly realistic model of the origin of the
 universe can be developed using the Friedmann equation from general
 relativity, elementary quantum mechanics, and the experimental values
 of $\hbar, c, G$ and the proton mass $m_p$.  The model assumes there are $N$
 space dimensions (with $N > 6$), and the potential constraining the
 radius $r$ of the invisible $N - 3$ compact dimensions varies as $r^4$.  In
 this model, the universe has zero total energy and is created from
 nothing.  There is no initial singularity.  If space-time is eleven
 dimensional, as required by M-theory, the scalar field corresponding
 to the size of the compact dimensions inflates the universe by about
 26 orders of magnitude (60 $e$-folds).  If $H_0 = 65$ \ksM, the
 energy density of the scalar field after inflation results in
 $\Om_\Lam = 0.68$,
 in agreement with recent astrophysical observations.  
\end{abstract}
\end{titlepage}

\renewcommand{\thefootnote}{\arabic{footnote}}
\setcounter{footnote}{0}
%

\section{Introduction}
           Analysis of cosmic microwave and X-ray background
radiation, extra-galactic radio sources, and Lyman-$\alpha$ lines from neutral
hydrogen in the universe indicates the three large-scale space
dimensions of our universe are isotropic and homogeneous \cite{Lahav}.  The
relevant variable in a homogeneous isotropic universe is the scale
factor \cite{Islam}.  For a closed universe, this scale factor is the radius of
curvature of the universe. 

Aleksandr Friedmann used Einstein's general relativity to obtain the
Friedmann equation for the scale factor $R$ of a homogeneous isotropic
universe \cite{Islam}:
\begin{equation}
\pone{\dd{R}{t}}^2 - \pone{{8\pi G\over 3}}\ve\pone{{R\over c}}^2= 
-kc^2
\label{eq:FR}
\end{equation}
where $\ve$  is the energy density of the universe, the gravitational
constant $G = 6.67 \times 10^{-8}$ cm$^3$/g sec$^2$, and $c = 3 \times
10^{10}$ cm/sec.	 For a 
closed universe, $k =1$; for a flat universe, $k=0$; and for an open
universe, $k=-1$.  At present, the energy density is \cite{MTW}
\[
\ve = \ve_r \pone{{R_0\over R}}^4 + \ve_m \pone{{R_0\over R}}^3 + \ve_\phi,
\]
 where $\ve_r, \ve_m$, and $\ve_\phi$  are, respectively, today's
values of the radiation, matter and 
scalar field energy densities, and $R_0$ is the scale factor of the
universe today.  Multiplied by ${1\over 2} m$, equation \eqr{eq:FR}
describes the 
motion of a fictitious particle with mass $m$ and energy $-{1\over 2}
k m c^2$ in the potential 
\[
V_R = - {m\over 2} \pone{{8\pi G\over 3}} \ptwo{\ve_r \pone{{R_0\over
R}}^4 + \ve_m \pone{{R_0\over R}}^3 + \ve_\phi} \pone{{R\over c}}^2.
\]
Since $\ve_r \approx 10^{-34}$ g/cm$^3$ $c^2$, 
$\ve_m \approx 10^{-29}$ g/cm$^3$ $c^2$, and  $R_0\approx 10^{28}$ cm,
the radiation term in the potential is approximately $-5.6 {m\over 2}
{10^{71}\over R^2}$ g (cm/sec)$^2$ and 
the matter term in the potential is about  $-5.6 {m\over 2}
{10^{48}\over R}$.  In the early universe,
when $R \ll 10^{-5} R_0$, the radiation term dominated, so this paper
neglects the matter term. 

According to the Friedmann equation, the universe began with the ``Big
Bang,'' when $R = 0$ and the energy density was infinite.  However, the
occurrence of infinities in a physical theory can signal a breakdown
in the theory \cite{GN}.  Because general relativity is not believed
to be valid at distances less than the Planck length, it has been
hoped that a quantum mechanical approach would avoid the problem of
the initial singularity in the Friedmann universe.

Besides the initial singularity, there are other problems with many
cosmologies based on the Friedmann equation.  For example, the
original Friedmann cosmology can't explain why the universe 
\begin{itemize}
\item is nearly flat (or, in other words, why it is so large that it
appears to be nearly flat),
\item  is so homogeneous, in that regions that could never have
communicated with each other by signals travelling at the speed of
light have the same matter/energy distribution, and 
\item has such a smooth distribution of matter.
\end{itemize} 
These problems (the flatness, horizon and smoothness problems)
\cite{Islam} are solved in inflationary models, where the universe
contains a scalar field.  Then, the universe can expand faster than
the speed of light in an exponential expansion that makes the universe
nearly flat, smooth and homogeneous.  Ideally, a quantum mechanical
approach to cosmology should also describe this inflationary phase.   

The usual approach to quantum cosmology, involving the Wheeler-De Witt
equation, is fraught with difficulties \cite{BS}.  However, the
canonical Hamiltonian quantization of standard cosmology by Elbaz {\em
et al} \cite{ENSMK} and Novello {\em et al} \cite{NSMK} avoids the
complexities 
of the Wheeler-De Witt equation.  For example, they show that the
quantum dynamics of a closed homogeneous and isotropic
radiation-dominated universe is equivalent to the quantum dynamics of
a particle moving in one dimension in a potential 
\[
V(q) = - {b^2\over 4 q^2},
\]
where $q$  is proportional to the radius of curvature of the universe
and $b$ is a  constant. 

The Schr\"odinger equation corresponding to equation \eqr{eq:FR}
(which is the same as the Schr\"odinger equation of Elbaz {\em et al}
\cite{ENSMK} and Novello {\em et al}
\cite{NSMK}) for a closed radiation-dominated Friedmann universe)
can be written as
\begin{equation}
- {\hbar^2\over 2m}\, \dd{^2\,}{R^2}\psi\, -\, {4\pi\, m G\over 3 c^2}\,
{\ve_r R_0^4\over R^2}\, \psi\  =\   
- {\hbar^2\over 2m}\, \dd{^2\,}{R^2}\psi
\, -\, {m G\over 2}\, {A\over R^2}\, \psi\  =\  - {m c^2\over 2} \psi
\label{eq:schrodinger}
\end{equation}
The solution to equation \eqr{eq:schrodinger} can be expressed as
\[
\psi = C_1\, R\, h_{ip - {1\over2}} (i\gk R),
\]
where $h$ is the spherical Hankel function \cite{MF}, or 
\[
\psi = C_2\, \gk\, \sqrt{R}\, K_{ip}(\gk R),
\]
 where $K$  is the modified Bessel function of the second kind
\cite{NSMK}.  
Here, $\gk=mc/\hbar$, the dimensionless index of the Bessel
 or Hankel function $p=\sqrt{{m^2GA\over\hbar^2}-{1\over4}}$, and
$C_1$ and $C_2$ are normalization constants.  This stationary state
wave function is zero at $R=0$.

          A time-dependent wave function describing the evolution of a
	closed radiation-dominated Friedmann universe must be a wave
	packet centered on the ``effective energy''  $-{1\over2} m c^2$.  Any
	superposition of stationary state solutions to equation
	\eqr{eq:schrodinger}  is
	also zero at $R=0$.  If general relativity were assumed valid at
	distances less than the Planck length, the Schr\"odinger
	equation of Elbaz {\em et al} would be valid at these distances.
	So, quantum mechanics forbids initial (and final)
	singularities in a closed radiation-dominated Friedmann
	universe.  Since the eigensolutions of the Schr\"odinger
	equations for flat and open radiation-dominated Friedmann
	universes are also zero at $R=0$, this conclusion can be extended
	to flat and open radiation-dominated Friedmann universes.
	This analysis indicates the need for a model of the universe
	where the radius of curvature is never zero.  One possibility
	is a quantum mechanical model allowing the universe to begin
	with a non-zero radius, by a quantum fluctuation from
	nothing. 

	Theories that adequately describe the four forces governing
the universe (gravity, electromagnetism, the weak force, and the
strong nuclear force) seem to require more than four space-time
dimensions.  If the extra dimensions are compact (for example, curled
up so tightly that their characteristic size is the Planck length
$10^{-33}$ cm), they are invisible under ordinary circumstances.  These
extra dimensions provide another degree of freedom allowing
development of the simple model \cite{Mongan1, Mongan2} for the origin
of the universe discussed in this paper.  In addition, the size of the
compact dimensions is a scalar field in the ordinary three dimensional
space described by the Friedmann equation, and this scalar field
produces the inflation needed to solve the problems of the earlier Big
Bang cosmologies.

\section{Overview of the Model}
\underline{Assumptions necessary to develop the model}
\begin{itemize}
\item The evolution of a homogeneous universe with $N$ space dimensions
can be described by a Schr\"odinger wave equation in an abstract 
$N$-dimensional Euclidean curvature space.  The coordinate of the universe
in each of the $N$ curvature space dimensions is the radius of curvature
of the corresponding dimension in the $N$-dimensional homogeneous
physical space.
\item The universe arises by a quantum fluctuation from nothing, so
all total quantum numbers in the curvature space (and in the physical
space) must be zero.
\item The potential in the curvature space describing our universe has
two terms.  One term involves only the radius of curvature of the
three space dimensions of a closed Friedmann universe.  The other term
involves only the characteristic radius of curvature of the $N-3$
compact dimensions.  So, the $N$-dimensional Schr\"odinger equation
separates into two equations describing two distinct subspaces of the
$N$-dimensional curvature space.  One subspace corresponds to our
homogeneous and isotropic Friedmann universe, and one subspace
corresponds to the compact dimensions.  The two subspaces do not
exchange energy, except when the curvature energy of the compact
dimensions drops to the ground state, injecting energy and entropy
into the Friedmann dimensions and causing inflation
\item The curvature of the Friedmann space is described by the quantum
mechanical analogue of equation \eqr{eq:FR}.  This is the s-wave Schr\"odinger
equation justified by Elbaz {\em et al} \cite{ENSMK} and Novello {\em et al}
\cite{NSMK}.  Since the total angular momentum in curvature space must
be zero, an s-wave Schr\"odinger equation must also be used to describe
the evolution of the compact space. 
\item In the initial state of the universe, the radius of curvature of
all of the dimensions was some small multiple $a$ of the Planck length,
and it was not changing.  In this initial state, gravity and the
strong-electroweak force had the same strength, and the Planck length
was about nineteen orders of magnitude larger than it is today.
Today, the characteristic radius of curvature of the compact
dimensions is $a$ times the Planck length. 
\end{itemize}

\noindent
\underline{Inputs needed for numerical estimates}
\begin{itemize}
\item The number of space-time dimensions: M-theory seems to be the
leading candidate for the fundamental theory of the four forces
governing the universe, so this paper assumes space-time is eleven
dimensional and $N = 10$.  
\item The long-range behavior of the effective potential constraining
the size of the compact dimensions: One can assume $V_r = k_n r^4$ to
get a universe similar to our own \cite{Mongan1, Mongan2}.  Or, one
can assume the energy 
spectrum of the compact dimensions must be that necessary for the
Tseytlin-Vafa \cite{TV} dimensional collapse scenario in string
theory.  In the latter case, it can be shown that the WKB
approximation for the energy spectrum requires $V_r = k_n r^4$.
\item The experimental values of four fundamental physical constants:
the proton mass $m_p = 1.67 \times 10^{-24}$ g\,, $\hbar= 1.05 \times
10^{-27}$  g cm$^2$/sec, $c$\,, and $G$.
\end{itemize}

\section{Model Development}

If the universe was created by a quantum fluctuation from nothing, it
must be closed, with all quantum numbers (including the total energy)
equal to zero.  If it is homogeneous, the relevant variables are the
radii of curvature of each of the dimensions. Consequently, an 
$(N + 1)$-dimensional quantum mechanical model of the origin of the
universe can 
be developed \cite{Mongan1,Mongan2}, where space is initially an
$N$-dimensional 
sphere.  The model is formulated in an $N$-dimensional Euclidean
curvature space (with $N > 6$) describing the curvature of a homogeneous
$N$-dimensional physical space.  In the simplest case, today's curvature
space has subspaces related to the Friedmann universe and the
$N - 3$ compact dimensions, and the curvature of all the compact
dimensions is the same.  The coordinate in each dimension of a state
in the curvature space is the radius of curvature of the corresponding
dimension of that state in the $N$-dimensional physical space.   

 When the total energy and total angular momentum in curvature space
 are zero, the Schr\"odinger equation for the $N$-dimensional radius of
 curvature is 
\[
-{\hbar^2\over 2m} \nabla^2_{\mbox{\tiny$\Re$}} \Psi +
 V_{\mbox{\tiny$\Re$}} \Psi = 0, 
\]
where $\Re$ is the magnitude of an $N$-dimensional vector $\vec{\bf
\Re}$ and $m$ is an effective mass. Today, the ``gravitational structure
constant''  ${Gm_p^2\over \hbar c} = 5.91 \times 10^{-39}$  is the ratio of
the strength of gravity to 
the strength of the strong force, the Planck mass  
$M=\sqrt{{\hbar c\over G}} = 2.18 \times 10^{-5}$ g, and the Planck length
$\gd = \sqrt{{\hbar G\over c^3}} = 1.62 \times 10^{-33}$ cm.  Initially,
gravity and the strong-electro-weak 
(SEW) force had equal strength, and ${G_i m_p^2\over \hbar c}=1$.  The
gravitational constant was 
initially $G_i = \pone{{M\over m_p}}^2 G= 1.70 \times 10^{38} G$, so ${G_i
m_p^2\over \hbar c} = {GM^2\over \hbar c} = 1$.  The Planck length was
$\gd_i=\sqrt{{\hbar G_i\over c^3}} = \pone{{M\over m_p}}\gd= 2.11 \times
10^{-14}$ cm and the Planck mass was $M_i=\sqrt{{\hbar c\over
G_i}}=m_p$. 
          
In this model, the universe can be envisioned to begin by a quantum
fluctuation from nothing into a spherically symmetric $N$-dimensional
universe with zero total energy, $\sqrt{\ev{\Re^2}}=a\gd_i$ and
$\dot\Re=0$.  The fundamental length in 
this curved space is the circumference of the $N$-sphere.  If
$a={1\over 2\pi}$, this
fundamental length is $\gd_i$.  In this initial state, none of the $N$ space
dimensions were distinguishable, and the only length scale is $\gd_i$ because
all forces were initially equal.  Such an initial state might be an
unbroken symmetry state of some fundamental theory of the four forces
governing the universe.

 After the symmetric initial state arose from nothing by a
 quantum fluctuation, a quantum tunneling transition occurred
 from the initial state to another state with zero total
 curvature energy, where $\Re^2=R^2+r^2$, $R$ is the radial
 coordinate in the 
 three dimensional subspace describing the curvature of the
 isotropic Friedmann universe, and $r$ is the radial coordinate
 in the $n = N - 3$ dimensional subspace describing the
 curvature of the compact dimensions.  At the transition,
 $\sqrt{\ev{R^2}}=\sqrt{\ev{r^2}} = a\gd_i$
 and $\dot{R}=\dot{r}=0$. This post-transition state was the
 beginning of 
 today's universe, where the size of the compact dimensions
 corresponds to a gauge singlet scalar field  $\phi$ \cite{Turner}
 that is constant throughout the Friedmann universe and
 drives inflation.

 Notice that the universe could arise directly from a quantum
 fluctuation from nothing into the post-transition stage
 described above.  However, the assumed transition from a
 completely symmetrical initial state allows a connection to
 more fundamental theories of the forces controlling the
 universe.

 The precise relation between the gauge singlet scalar field
 $\phi$ and the characteristic size of the compact dimensions
 is not important in this simple model.  One realization of
 the model might involve a scalar field related to the size
 of the compact dimensions by $\phi={\zeta\over a \gd_i}
 \sqrt{{\hbar \over c}}\ln\pone{{a\gd_i\over r}}$.   Then,
 the scalar field $\phi=0$
 when $r=a\gd_i$.  The value of the real number  $\zeta$
 would have to be 
 obtained from a more fundamental theory of the four forces
 governing the universe, such as M-theory. 

 The model assumes $V_{\mbox{\tiny$\Re$}}= V_R+ V_r$, so $\Psi =
\Psi(R)\Psi(r)$ and
\[
\ptwo{{1\over \Psi(R)} {-\hbar^2\over 2m}\nabla^2_R \Psi(R) + V_R}
+ \ptwo{{1\over \Psi(r)}{-\hbar^2\over2m}\nabla^2_r \Psi(r) + V_r} = 0,
\] 
where each bracket is a constant, denoted $-E$ and $E$ respectively.  So, the
curvature energy of the closed Friedmann universe is less than zero,
the curvature energy of the extra dimensions is greater than zero, and
the total curvature energy of the universe is zero.  In this model,
the universe evolved from an excited state with large curvature
energies $-E$ and $E$  (where $|E|\gg {1\over 2} m c^2$) reached by a
quantum transition from 
the spherically symmetric, zero curvature energy initial state with
indistinguishable space dimensions.  Today's universe is a quantum
state where the curvature energy of the Friedmann universe has the
Einstein value $-{1\over 2} m c^2$, and the compact dimensions are in
the ground 
state of the potential $V_r$, with curvature energy ${1\over 2} m c^2$.  The
resulting simple quantum mechanical model for the origin of the
universe is schematically outlined in Figure~1 (the details of
Figure~1 will be explained below).  The model provides a quantum
theory of 
space, but it is certainly not the long-sought quantum field theory of
gravity that will truly unify quantum mechanics and general
relativity.

\begin{figure}[h]
\centerline{\begin{picture}(0,0)%
\epsfbox{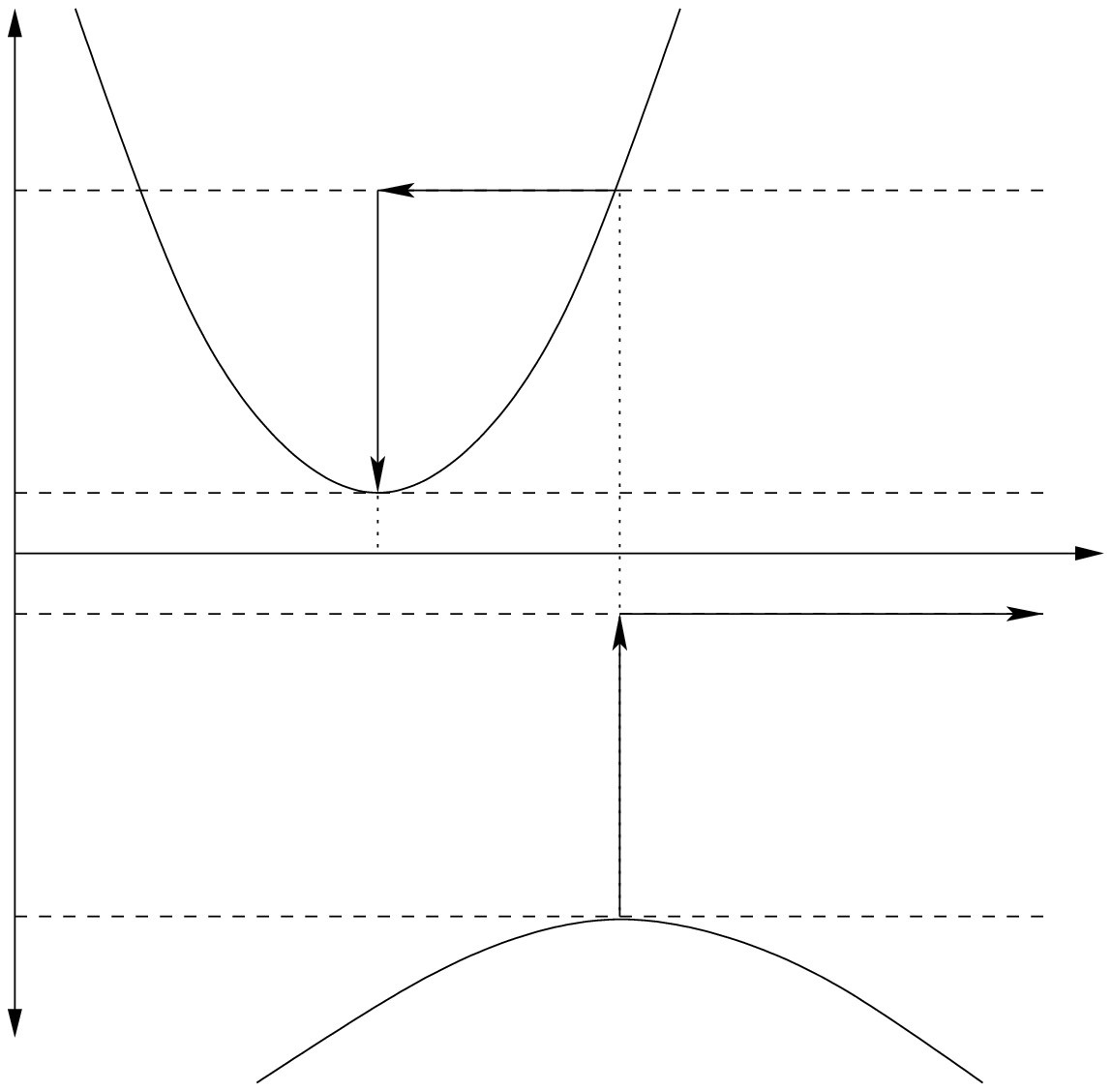}%
\end{picture}%
\setlength{\unitlength}{3947sp}%
\begin{picture}(6825,5758)(1126,-6689)
\put(7951,-4036){\makebox(0,0)[lb]{\smash{\SetFigFont{14}{16.8}
{\familydefault}{\mddefault}{\updefault}{$r, R$}%
}}}
\put(2000,-2266){\makebox(0,0)[lb]{\smash{\SetFigFont{12}{14.4}
{\familydefault}{\mddefault}{\updefault}{$E'$}%
}}}
\put(5851,-1606){\makebox(0,0)[lb]{\smash{\SetFigFont{12}{14.4}
{\familydefault}{\mddefault}{\updefault}{$V_r\approx kr^4$}%
}}}
\put(4096,-4186){\makebox(0,0)[lb]{\smash{\SetFigFont{12}{14.4}
{\familydefault}{\mddefault}{\updefault}{$a\delta$}%
}}}
\put(5476,-4186){\makebox(0,0)[lb]{\smash{\SetFigFont{12}{14.4}
{\familydefault}{\mddefault}{\updefault}{$a\delta_i$}%
}}}
\put(4126,-6631){\makebox(0,0)[lb]{\smash{\SetFigFont{12}{14.4}
{\familydefault}{\mddefault}{\updefault}{$V_R=-{4\pi\delta_\phi\over
3}\left( {A'\over R^2} + \varepsilon_\phi R^2\right)$}%
}}}
\put(2291,-1111){\makebox(0,0)[lb]{\smash{\SetFigFont{14}{16.8}
{\familydefault}{\mddefault}{\updefault}{$E$}%
}}}
\put(1046,-4261){\makebox(0,0)[lb]{\smash{\SetFigFont{12}{14.4}
{\familydefault}{\mddefault}{\updefault}
{$-{1\over 2}\left(\beta_n\over f_n a\right) M c^2$}%
}}}
\put(1876,-5836){\makebox(0,0)[lb]{\smash{\SetFigFont{12}{14.4}
{\familydefault}{\mddefault}{\updefault}{$-E'$}%
}}}
\put(1201,-3661){\makebox(0,0)[lb]{\smash{\SetFigFont{12}{14.4}
{\familydefault}{\mddefault}{\updefault}
{${1\over 2}\left(\beta_n\over f_n a\right) M c^2$}%
}}}
\put(6076,-3061){\makebox(0,0)[lb]{\smash{\SetFigFont{12}{14.4}
{\familydefault}{\mddefault}{\updefault}{Compact Dimensions}%
}}}
\put(6001,-5311){\makebox(0,0)[lb]{\smash{\SetFigFont{12}{14.4}
{\familydefault}{\mddefault}{\updefault}{Friedmann Dimensions}%
}}}
\put(6451,-4636){\makebox(0,0)[lb]{\smash{\SetFigFont{12}{14.4}
{\familydefault}{\mddefault}{\updefault}{Inflation}%
}}}
\end{picture}
}
\caption{Schematic sketch of the initiation of inflation.  
(Not drawn to scale).  Here, $E'={(n-1)(n-3)M c^2\over 16 \beta_n f_n^5
a}\pone{\delta_i\over \delta}^4$.}
\end{figure}

 In the model, the Schr\"odinger equation for $\psi(R)$ is the quantum
  analog of the Friedmann equation for a universe containing radiation
  and a scalar field: 
\begin{eqnarray}
-{\hbar^2\over 2m_\phi}\, \dd{^2\ }{R^2} \psi\, -\, {4\pi\, m_\phi
  G_\phi\over 3}\, (\ve_r + \ve_\phi) \pone{{R\over c}}^2 \psi &=& -
  E'\psi \nonumber \ \ \ \mbox{ or} \\
-{\hbar^2\over 2m_\phi}\, \dd{^2\ }{R^2} \psi\, -\, {4\pi \gd_\phi\over 3}
\pone{{A'\over R^2} + \ve_\phi R^2}\psi &=&-E'\psi \label{eq:psiR}
\end{eqnarray}
where the scalar field energy density
$\ve_\phi=\dot{\phi}^2+V_\phi(\phi)$ models the effect of the compact
dimensions on the Friedmann universe.   The subscript $\phi$ indicates that
$G$, the effective mass, the Planck length, and the scalar field energy
density depend on the value of the scalar field $\phi$.  If
$R\rightarrow 0$  or $\ve_r\gg\ve_\phi$, equation
\eqr{eq:psiR} reduces to the Schr\"odinger equation \eqr{eq:schrodinger}
for a radiation dominated
universe \cite{ENSMK,NSMK}.  An s-wave Schr\"odinger equation must be
used for the 
compact dimensions to make the total N-dimensional ``angular momentum''
in the curvature space zero.  Writing $\Psi=R^{-1}\psi(R)
r^{-(n-1)/2}\psi'(r)$, the separated Schr\"odinger
equation becomes 
\[
\ptwo{{1\over \psi(R)}{-\hbar^2\over 2m} \dd{^2\psi(R)}{R^2} + V_R}
+
\ptwo{{1\over\psi'(r)}{-\hbar^2\over 2m} \dd{^2\psi'(r)}{r^2}+
\pone{{\hbar^2(n-1)(n-3)\over 8mr^2} + V_r}} = 0 .
\]
If $V_r = k_n r^4$, the model produces a universe like our own
\cite{MF,Mongan1}  if the minimum of the effective potential
${\hbar^2(n-1)(n-3)\over 8 m r^2} + k_n r^4$ in the
compact dimensions is near $r=a\gd$.  Specifically, if the minimum of the
effective potential in the compact dimensions is at $r=f_n a\gd$,
$k_n={\hbar^2(n-1)(n-3)\over 16 m f_n^6 a^6 \gd^6}$.  The factor
$f_n$, calculated below, ensures that the radius of the compact 
dimensions today is $\sqrt{\ev{r^2}}=a\gd$.  Approximating the
effective potential for the 
compact dimensions by a harmonic oscillator potential near its
minimum, the ground state energy of the compact dimensions is
$E_{\mbox{\scriptsize g}}={\hbar^2\beta_n^2\over 2 m f_n^2 a^2 \gd^2}$, where
$\beta_n^2 = {3\over 8}(n-1)(n-3)+ \sqrt{{3\over2}(n-1)(n-3)}$.
  Setting $E_{\mbox{\scriptsize g}} = {1\over2} m c^2$ establishes the effective mass as
$m=\pone{{\beta_n\over f_n a}}M$,
so $k_n={\hbar^2(n-1)(n-3)\over 16 \beta_n M f_n^5 a^5 \gd^6} =
{(n-1)(n-3) M c^2\over 16 \beta_n f_n^5 a^5 \gd^4}$. 
The ground state wavefunction of the compact dimensions is 
\[
\psi'(r) \approx \pone{{2\over \pi}}^{1\over 4} \sqrt{{\gamma_n\over f_n a
\gd}} \exp\pone{-{\gamma_n^2(r-f_na\gd)^2\over f_n^2a^2\gd^2}},
\]
where $\gamma_n^2=\sqrt{{3(n-1)(n-3)\over 8}}$.
Setting $x=r-f_n a\delta$,  
\begin{eqnarray}
\ev{r^2} &\approx&
\pthree{\int^0_{-f_n a \gd} (x+f_n a \gd)^2 \exp\ptwo{-\pone{{\gamma_n x
\over f_n a\gd}}^2} {\mbox d}x +  
\int_0^\infty  (x+f_n a \gd)^2 \exp\ptwo{-\pone{{\gamma_n x \over f_n a
\gd}}^2} {\mbox d}x } \nonumber\\
&& \times \pthree{
\int_{-f_n a\gd}^0  \exp\ptwo{-\pone{{\gamma_n x \over f_n a \gd}}^2} {\mbox
d}x 
+ \int_0^\infty  \exp\ptwo{-\pone{{\gamma_n x \over f_n a \gd}}^2} {\mbox d}x
}^{-1} \nonumber \\
&=&
{ f_n^2 a^2\gd^2 \pthree{-{\exp(-\gamma_n^2)\over \gamma_n} + \sqrt{\pi}
\ptwo{1+ \mbox{Erf}(\gamma_n)}\pone{1+{1\over2\gamma_n^2}}}
\over
\pthree{\sqrt{\pi}\ptwo{1+\mbox{Erf}(\gamma_n)}}}, \nonumber
\end{eqnarray}
where Erf$(\gamma_n)$  is the error function of $\gamma_n$.  So, 
$\sqrt{\ev{r^2}}=a\gd$ if
\[
f_n={ \pthree{-{\exp(-\gamma_n^2)\over \gamma_n} + \sqrt{\pi}
\ptwo{1+ \mbox{Erf}(\gamma_n)}\pone{1+{1\over2\gamma_n^2}}
\over
\sqrt{\pi}\ptwo{1+\mbox{Erf}(\gamma_n)}}}^{-{1\over2}}.
\]

 After the quantum transition from the initial state, and just prior
 to inflation, the universe was still in a symmetric state with 
 radius $a\gd_i$ , as indicated in Figure 1.  Immediately after the
 transition, the compact dimensions were in a highly excited state of
 the effective potential $V_r$, with wave packet localized at the
 classical turning radius $r=a\gd_i$, and curvature energy  
 $E'={(n-1)(n-3)Mc^2\over 16 \beta_n f_n^5 a}\pone{{\gd_i\over\gd}}^4$.
  The curvature
 energy in the Friedmann dimensions at transition, $-E'$, coincided with
 the top of the effective potential in equation \eqr{eq:psiR}, at
 $R_{\mbox{\scriptsize peak}} = \gd_i$, where $R_{\mbox{\scriptsize
 peak}}^{\,4}  = {A'\over \ve_\phi}$.  So
 the transition resulted in a state with wave packet centered at $R=a\gd_i$, in
 unstable equilibrium, with $\dot{R}=0$.  At transition $\ve_\phi =
 {A'\over a^4\gd_i^4}$, so $\ve_r=\ve_\phi$ and
  ${8\pi m_\phi G_\phi A'\over 3 c^2 a^2 \gd_i^2} =
 {(n-1)(n-3)Mc^2\over 16 \beta_n f_n^5 a}\pone{{\gd_i\over\gd}}^4$.
  Incidentally, the effective $r^3$ force constraining the size of the
 compact dimensions in this model is related to the effective $1/R^3$
 radiation force in the Friedmann universe by the replacement
 $R\rightarrow 2^{-2/3}a^2\gd_i^2/r$. 
	
When the curvature energy of the compact dimensions dropped to the
 ground state energy ${1\over2}\pone{{\beta_n\over f_n a}Mc^2}$, the
 curvature energy of the Friedmann 
 universe was raised to the Einstein value
 $-{1\over2}\pone{{\beta_n\over f_n a}Mc^2}$.  The scalar 
 field $\phi$ changed from its initial value $\phi_i$ to its present
 value $\phi_f$
 as the characteristic size of the compact dimensions
 decreased from $\gd_i$ to $\gd$, $G$ decreased from $G_i$ to its
 present value, 
 and the Planck mass increased from $m_p$ to its present value
 $M=\sqrt{{\hbar c\over G}}= 2.18\times 10^{-5}$ g.

In this model, inflation occurred when the characteristic size of the
compact dimensions shrank from $\gd_i$ to $\gd$  and the curvature
energy of the compact dimensions dropped from the transition energy  
$E'$ to the ground state energy $E_{\mbox{\scriptsize g}}$, raising
the curvature energy of 
the Friedmann dimensions to $-E_{\mbox{\scriptsize g}}$ .  Entropy was
injected into the 
scalar field in the Friedmann dimensions, and transferred to radiation
as the scalar field decayed during inflation.  When
$R>R_{\mbox{\scriptsize peak}}=a\gd_i$, the $\ve_\phi$  term in
equation \eqr{eq:psiR} dominated and the Friedmann universe inflated.
Figure 1 schematically indicates the process that initiated inflation. 

Since $\dot{r}=0$ at transition, $\dot{\phi}=0$  and the scalar field
 energy density at transition was
 $\ve_\phi=\dot{\phi}^2+V_\phi(\phi_i)=V_\phi(\phi_i)$ .  When the
 compact dimensions reached their ground state at the end of
 inflation, $\ev{\dot{r}^2}=\ev{\dot{\phi}^2}=0$ thereafter, and the
 scalar field energy density remained 
 constant at $\ve_\phi=\dot{\phi}^2+V_\phi(\phi_f)=V_\phi(\phi_f)$.
 This is consistent with the observation that, after 
 the scalar field energy stopped decaying to radiation at the end of
 inflation, the energy conservation equation can be separated into two
 parts, one for radiation and one for the scalar field \cite[pg. 727]{MTW}.
 Then the scalar field pressure is
 $p_\phi=\dot{\phi}^2-V_\phi(\phi)=-V_\phi(\phi_f)=-\ve_\phi$, and the
 scalar field energy 
 density is constant thereafter.  So, at the end of inflation, after
 the scalar field stopped decaying to radiation, $\ve_\phi \ll
 \ve_{\mbox{\scriptsize rad}}$, and the radiation-dominated universe
 satisfied equation \eqr{eq:schrodinger} with 
\begin{equation}
A={(n-1)(n-3)a^2\over 8 \beta_n^2 f_n^4} {\hbar\over c}
\pone{{\gd_i\over\gd}}^6. \label{eq:A}
\end{equation}

\section{Numerical Estimates}

M-theory (involving eleven-dimensional space-time) seems to be the
 leading candidate for the theory of the four forces governing
 the universe, so the remainder of this paper assumes $N= 10$,
 $n = 7$ and $a =1/2\pi$.  Then, $\beta_7^2  = 15$, $\gamma_7=3$,
 and $f_7 = 0.93$.  From equations \eqr{eq:schrodinger} and
 \eqr{eq:A}, $V_{\mbox{\scriptsize radiation}} = 7.8
 {m\over2}{10^{67}\over R^2}$ g cm$^2$/sec$^2$, and the calculated
 radiation density is $\ve_r= 1.4 \times 10^{-38}$ g/cm$^3$c$^2$ if
 $R_0 = 10^{28}$ cm. 
 However, the model does not explicitly account for strong-electroweak
 symmetry breaking.  Nuclear energy levels, determined by the strong
 force, are roughly $10^6$ times the atomic energy levels determined by
 the electromagnetic force, indicating that strong-electroweak
 symmetry breaking would increase the initial radiation density of the
 universe by up to six orders of magnitude.  When some of this
 primordial radiation decays to hadrons, the remaining photons could
 then result in today's microwave background radiation energy density
 of $\ve_r = 4.66 \times 10^{-34}$ g/cm$^3$c$^2$ \cite{PDG}. 

 The extent of inflation can be estimated by assuming the curvature
energy of the compact dimensions dropped instantaneously to the ground
state energy when the size of the compact dimensions reached
$\ev{r}=a\gd$.  The ratio of ${E'\over E_{\mbox{\tiny g}}}$ the
transition energy to 
the ground state energy is ${E'\over E_{\mbox{\tiny g}}}
={(n-1)(n-3)\over 8\beta_n^2 f_n^4}\pone{{\gd_i\over\gd}}^4= 7.73
\times 10^{75}$.  At the end of 
inflation, the compact 
dimensions were in their ground state and could no longer transfer
energy to the Friedmann dimensions.  The scalar field had decayed to
radiation and could no longer transfer entropy to the radiation field.
Assuming the curvature energy of the compact dimensions dropped
instantaneously to the ground state when the size of the compact
dimensions reached $\ev{r}=a\gd$ \cite{Mongan2}, the entropy of the compact
dimensions was reduced by a factor of $0.13 \times 10^{-75}$ and the entropy of
the scalar field in the Friedmann universe was increased by a factor
of $7.73 \times 10^{75}$ at the beginning of inflation. Then, if $T_0$ was the
temperature of the Friedmann universe at the end of inflation, when
entropy in the scalar field stopped being transferred to radiation,
the temperature of the scalar field at the beginning of inflation was
$(7.73 \times 10^{75})^{1/3} T_0 = 1.98 \times 10^{25} T_0$.  During
isentropic expansion, 
$(RT)^3$ remains constant.  So, the injection of entropy from the
collapse of the compact dimensions increased the scale factor of the
Friedmann universe by a factor of $1.98 \times 10^{25}$ as the Friedmann
universe expanded exponentially and isentropically (driven by the
scalar field), until inflation ended when $\dot{\phi}\approx0$ and the
temperature of the 
universe was $T_0$.  This 58 $e$-fold inflation is within the limits set by
fluctuations in the microwave background radiation \cite{Peacock}.  If
the strong-electroweak symmetry broke during inflation, as it must
have to prevent monopole dominance, the temperature should increase by
a factor of about $(10^6)^{1/3}$ at some instant during inflation.  This
will increase the inflation by a factor of about $(10^6)^{1/3} =  e^{4.6}$, for
a total inflation of more than 60 $e$-folds. 

In this model, the vacuum energy density today is the scalar field
energy density per unit coordinate volume (i.e., the vacuum energy
density is the scalar field energy density per unit coordinate volume
multiplied by the number of unit coordinate volumes in the universe
and divided by the volume of the universe expressed in terms of unit
coordinate volumes).  The vacuum energy density has remained constant
since the scalar field decoupled from the radiation field at the end
of inflation.  It is the energy density associated with the creation
of 1 cm$^3$ of space during the continuing expansion of the universe. 

The scalar field energy density in the co-moving volume of the
universe at the beginning of inflation can be obtained from the
Schr\"odinger equation \eqr{eq:psiR}.  At the moment of transition, when
$\dot{R}=0$ and the scale factor $R=a\gd_i$, equation \eqr{eq:psiR}
becomes ${4\pi\gd_i\over3}2\ve_\phi a^2 \gd_i^2=E'$, because
$\ve_r=\ve_\phi$  at transition.  This scalar field energy density
$\ve_\phi$  is the total scalar field energy in the 3-sphere (with
scale factor $R=a\gd_i$) comprising the initial Friedman universe,
divided by the volume of the 3-sphere.  It is related to $\eta_\phi$,
the scalar field energy density per unit coordinate volume at the
beginning of inflation, by
$\ve_\phi={1\over2\pi^2a^3\gd_i^3}\eta_\phi$ \cite{Hu}.  So, the
energy density per 
unit coordinate volume of the scalar field in the Friedmann universe
at the start of inflation was  
\[
\eta_\phi =\pone{3\pi\over4 f_7^5} 2.18 \times 10^{92}\, \mbox{g
cm$^{-1}$ sec$^{-2}$} = 7.38 \times 10^{92} \, \mbox{g cm$^{-1}$ sec$^{-2}$}.
\]
A spatially constant scalar field has only one degree of freedom, and
the energy density in the scalar field is proportional to the fourth
power of the temperature.  So, the energy density per unit coordinate
volume of the scalar field at the end of inflation was 
\[
\eta_e = {T_0^4\over (1.98\times10^{25} T_0)^4} 
7.38 \times 10^{92} \, \mbox{g cm$^{-1}$ sec$^{-2}$}  = 4.80 \times
10^{-9}  \, \mbox{g cm$^{-1}$ sec$^{-2}$} = \ve_e. 
\]

Primack \cite{Primack} finds $H_0= 65$ km sec$^{-1}$ Mpc$^{-1}$, based
on recent astrophysical measurements, so the critical density 
$\rho_c={3H_0^2\over8\pi G}= 7.91 \times 10^{-30}$ g cm$^{-3}$ and the
critical energy density $\ve_c=\rho_c c^2 = 7.11 \times 10^{-9} $g cm$^2$
sec$^{-2}$ cm$^{-3}$. Therefore, in this model,
$\Om_\Lam={\ve_e\over\ve_c}= 0.68$, in agreement with the value
$\Om_\Lam\approx 0.7$
obtained by Primack \cite{Primack} from astrophysical data.  

\section{Conclusions}

M-theory, a prime candidate for the fundamental theory of the forces
 governing the universe, requires eleven dimensional
 space-time with seven compact space dimensions.  The model
 outlined above simulates two key features of M-theory as
 applied to cosmology: a scalar field in the Friedmann
 equation \cite{Muko}, and the Tseytlin-Vafa dimensional
 compactification scenario \cite{TV}.  The vacuum energy
 density in the model can be identified with the cosmological
 constant/``dark energy'' in the Friedmann equation for our
 four-dimensional universe that arises in M theory
 \cite{Muko}. 

The Schr\"odinger equation for the effective potential
${\hbar^2(n-1)(n-3)\over 8 m r^2} + k_n r^4$  in the compact dimensions
cannot be solved exactly.  However, the energy levels are related to
the classical turning radius $r_{\mbox{\tiny T}}$ by $E=k_n
r_{\mbox{\tiny T}}^4$ and the WKB 
approximation shows that allowed values of $r_{\mbox{\tiny T}}$
satisfy $r_{\mbox{\tiny T}}^3 
\propto \pone{j + {1\over 2}} \pi$ , where $j$ is an integer.  This indicates
that the energy spectrum varies as the radius of the compact
dimensions if the effective potential varies as $r^4$ for large $r$. 

The solutions to the Schr\"odinger equation \eqr{eq:schrodinger} for a
radiation dominated Friedmann universe involve the factor
 $p=\sqrt{{8\pi m^2 G \ve_r R_0^4\over 3\hbar^2 c^2} - {1\over 4}}$.  
 For $p^2\gg 1$, the energy levels \cite{MF}
 of a radiation dominated Friedmann universe are $E_n = {m c^2\over 2}
\exp\pone{{2\pi j'\over p}}\approx {m c^2\over 2}\pone{1+{2\pi j'\over
p}}$, where $j'$  is an integer.  The r.m.s. radius of the Friedmann
dimensions is 
\[
\sqrt{\ev{R^2}} = \ptwo{
{ \int_0^\infty \, R^3\,  K_{ip}(\gk R)\, K_{-ip}(\gk R)\, \mbox{d}R
\over
\int_0^\infty R\,  K_{ip}(\gk R)\, K_{-ip}(\gk R)\, \mbox{d}R}}^{1\over2}
= \gk^{-1} \sqrt{{2(1+p^2)\over3}}.
\]
Since $\gk^{-1}={\hbar\over mc} = {f_n a\over \beta_n}\gd$,
\[
\sqrt{\ev{R^2}}\ =\ {f_n a\over \beta_n}\, \gd\, \sqrt{2(1+p^2)\over3}
\ \approx\   p \pone{f_n a \gd \over \beta_n} \sqrt{2\over 3},
\]
so the energy spectrum goes as $1/R$.    

Tseytlin and Vafa \cite{TV} claim that string winding modes in $N - 3$
dimensions, with an energy spectrum varying as $r$, compact those
dimensions.  The constricting effect of winding modes can only be
overcome in three of the space dimensions, where the energy spectrum
of the momentum modes varies as $\gd_i/R$, according to the T-duality of string
theory.  In the expanding three dimensions, the massless string
momentum modes are the photons of the radiation-dominated Friedmann
universe. So, the model simulates the dimensional compactification
scenario in string theory envisioned by Tseytlin and Vafa.
Furthermore, once the number of space dimensions is specified, and 
$V_r= k_n r^4$ is chosen to accommodate the Tseytlin/Vafa dimensional
compactification mechanism, only the experimental values of $\hbar, c, G$ and
the proton mass $m_p$ are needed to obtain numerical results from the
model.

To conclude, a simple and surprisingly realistic $N + 1$ dimensional
quantum mechanical model of the universe can be developed using the
Friedmann equation from general relativity, elementary quantum
mechanics, and measured values of $\hbar , c, G$, and the proton mass. The
model suggests that the {\em details} of extra-dimensional collapse are less
important than the {\em fact} of extra-dimensional collapse in explaining
the inflation of our three-dimensional universe and the size of the
cosmological constant/vacuum energy density.

\end{document}